\documentclass[journal,twoside]{IEEEtran}
\usepackage{cite}
\usepackage{graphicx}
\usepackage{psfrag}
\usepackage{subfigure}
\usepackage{amsmath}
\usepackage{multirow}
\usepackage{array}
\usepackage{color}
\usepackage{booktabs}
\usepackage{threeparttable}
\usepackage{booktabs}
\usepackage{makecell}

\usepackage{amsthm}
\usepackage{mathrsfs}
\usepackage{lipsum}
\usepackage{tabularx}

\usepackage{epsfig}
\usepackage{stfloats}
\usepackage{enumerate}
\usepackage{times}
\usepackage{bm}
\usepackage{footnote}

\begin{document}
\title{{\LARGE{An Enhanced Access Reservation Protocol with a Partial Preamble Transmission Mechanism in NB-IoT Systems}}}
\author{Taehoon Kim, Dong Min Kim, Nuno Pratas, Petar Popovski, and Dan Keun Sung

\thanks{This work has been in part supported by the National Research Foundation of Korea (NRF) grant funded by the Korea government (MSIP) (No. 2014R1A2A2A01005192), and in part by the European Research Council (ERC Consolidator Grant Nr. 648382 WILLOW) within the Horizon 2020 Program.}
\thanks{T. Kim, and D. K. Sung are with the School of Electrical Engineering, KAIST, Daejeon 34134, Republic of Korea (e-mail:
 \{thkimm, dksung\}@kaist.ac.kr).}
\thanks{D. M. Kim, N. Pratas, and P. Popovski are with the Department of Electronic Systems, Aalborg University, Aalborg 9000, Denmark (e-mail:
	\{dmk, nup, petarp\}@es.aau.dk).}
 }

\maketitle

\begin{abstract}
In this letter, we propose an enhanced Access Reservation Protocol (ARP) with a partial preamble transmission (PPT) mechanism for the narrow band Internet of Things (NB-IoT) systems. The proposed ARP can enhance the ARP performance by mitigating the occurrence of preamble collisions, while being compatible with the conventional NB-IoT ARP. We provide an analytical model that captures the performance of the proposed ARP in terms of false alarm, mis-detection and collision probabilities. Moreover, we investigate a trade-off between the mis-detection and the collision probabilities, and optimize the proposed ARP according to the system loads. The results show that the proposed ARP outperforms the conventional NB-IoT ARP, in particular at heavier system loads.
\end{abstract}

\begin{IEEEkeywords}
NB-IoT, Random access, Preamble structure.
\end{IEEEkeywords}

\vspace{-0.2cm}
\section{Introduction}\label{CH1:Introduction}
The support for Internet of Things (IoT) plays a major role in the evolution of
wireless/cellular systems, including the upcoming 5G communication
systems~\cite{Palattella2016JSAC}. There are two general classes of IoT transmission
technologies: (1) Low-power wide-area (LPWA) communication technologies  (e.g.,
LoRa~\cite{lora}, Sigfox~\cite{sigfox}, IEEE 802.11ah) that operate in unlicensed
spectra, and offer low-cost devices and ease of network deployment; (2) Cellular IoT
technologies (e.g., LTE Cat-1, LTE Cat-0, LTE-M Cat-M1, NB-IoT) that operate in
licensed spectra. Among these, NB-IoT is a recent technology that has gained a
significant momentum, as observed by the fast standardization during
2016~\cite{TS36211} and the increasing number of deployments.

NB-IoT is designed to accommodate a massive number of low-throughput, low-cost, and delay-tolerant devices. Similarly to LTE networks, each device registers at the network through an Access Reservation Protocol (ARP), i.e., random access, in which time and frequency misalignments can be adjusted.
The preamble sequence is transmitted at the first step of the ARP, however, the preamble structure is no longer based on Zadoff-Chu sequence. The preamble design and detection algorithm for the NB-IoT ARP was presented in~\cite{Lin2016WCL}, while an overview of the NB-IoT air interface was given in~\cite{Wang2017CM}.

The performance of the ARP significantly degrades due to preamble mis-detections and collisions. The preambles in NB-IoT systems were designed with a goal of extending coverage and reducing the occurrence of mis-detections. Yet, the number of devices within cell coverage is expected to grow, leading to an increasing number of preamble collisions. This motivates us to configure the preamble structure by considering the effect of collisions.

This letter presents an enhanced ARP with a partial preamble transmission (PPT) mechanism that leverages the trade-off between mis-detections and collisions. We can significantly improve the performance of the ARP by puncturing the preamble sequence through the proposed PPT mechanism. It is worth noting that the proposed ARP requires only minor modifications on how the preambles are transmitted and detected, which can be easily implemented in NB-IoT systems.

\vspace{-0.2cm}
\section{Access Reservation Protocol in NB-IoT}\label{Backgrounds}
In NB-IoT systems, the ARP consists of 5 steps, and the detail of each step is described as follows:
\begin{itemize}
    \item \textbf{Step 1}: The device selects one of the available preamble sequences, and transmits it in the Narrowband Physical Random Access Channel (NPRACH);
    \item \textbf{Step 2}: The eNodeB detects the preamble sequences and responds to the detected preamble sequences by sending a Random Access Response (RAR), which includes the index of the preamble sequence, the time alignment (TA) offset and an uplink grant;
    \item \textbf{Step 3}: The device proceeds with the signaling information exchange
        on the resources indicated by the RAR, termed the \emph{RRC Connection Request};
    \item \textbf{Step 4}: The eNodeB acknowledges the signaling information received from the device with the \emph{RRC Connection Setup} message;
    \item \textbf{Step 5}: Finally, the device transmits its data concatenated with the \emph{RRC Connection Setup Complete} message.
\end{itemize}

\begin{figure}[t]
    \centering
    \includegraphics[width=9.0cm]{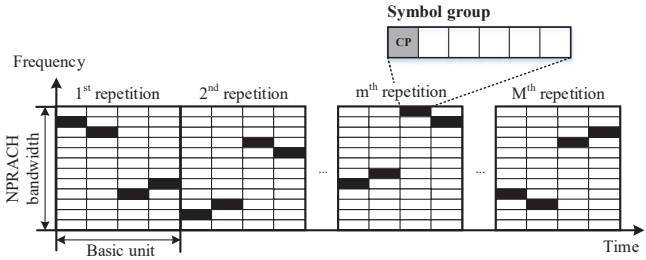} 
    \vspace{-0.3cm}
    \caption{Structure of a preamble sequence in a single NPRACH.}
    \label{fig:preamble_structure}
\vspace{-0.5cm}
\end{figure}

NB-IoT can be implemented with a $180$kHz bandwidth, composed of $48$ sub-carriers with each sub-carrier spacing of $3.75$kHz, which can be configured for the NPRACH; allowing 12, 24, 36, or 48 orthogonal RA preamble sequences to be available. A preamble sequence is composed of multiple \emph{symbol groups} as shown in Fig.~\ref{fig:preamble_structure}, where a single \emph{symbol group} consists of a cyclic prefix (CP) and $\xi$ symbols. The rationale behind this structure is to reduce the relative CP overhead, and, thus, $\xi$ should be carefully set to be sufficiently small so that the channel condition remains the same within each of the symbol groups. All of symbols have the same value of ``1''. $\nu$ symbol groups configure \emph{a basic unit} for repetition, which can be repeated $M$ times, where $M\!=\!2^q$, $q\!=\!0,...,7$. Therefore, the length of a preamble sequence, $L$, can be represented as $L\!=\!\nu\!\cdot\!M$. {$\xi$ and $\nu$ are commonly set to 5 and 4, respectively.}

Each symbol group uses a single carrier, and hops across frequency to facilitate to estimate uplink timing alignment at the eNodeB.
Thus, selecting a preamble sequence implies that each device selects a hopping pattern, and $\Omega(\cdot)$ represents a mapping function from a preamble index to the set of sub-carrier indices which are used by the corresponding preamble sequence. The sub-carrier index used by the $l$-th symbol group of the $i$-th preamble sequence is denoted as $\omega_l^i$, $l\!=\!1,...,L$, and, thus, $\Omega(i)\!=\!\left[\omega_1^i,...,\omega_L^i \right]$.
Finally, NB-IoT networks can support up to 3 coverage classes, which can be configured with different values of $M$, in order to support specific coverage requirements. However, in this letter, we focus on a single coverage class.

\section{Proposed Access Reservation Protocol with a Partial Preamble Transmission Mechanism}\label{Proposed_Scheme}
\subsection{Partial Preamble Transmission Mechanism}\label{PPT_mechanism}
The key idea of a partial preamble transmission (PPT) mechanism is to allow each device to transmit \emph{a fraction of a preamble sequence}, which is called \emph{a partial preamble sequence} (PPS). According to the PPS configuration, the NPRACH can be virtually divided into multiple sub-NPRACHs, {each of} which is named \emph{a partial unit}, where a single PPS is transmitted. We note that this approach does not alter the intrinsic structure of the protocol, and is instead a reconfiguration of the protocol.

\begin{figure}[t]
    \centering
    \subfigure[$L_p=4$ ($M_p=1$), $G=4$.]
    {\includegraphics[width=7.5cm]{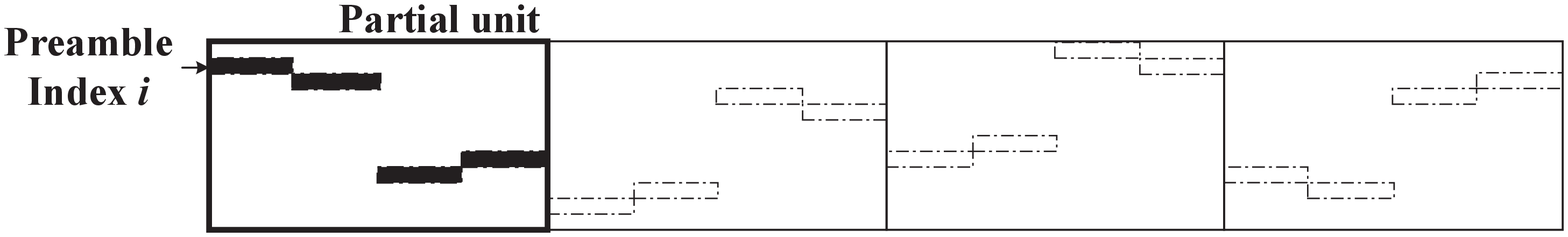}\label{fig:pt_1}}\\ 
    \vspace{-0.2cm}
    \subfigure[$L_p=8$ ($M_p=2$), $G=2$.]
    {\includegraphics[width=7.5cm]{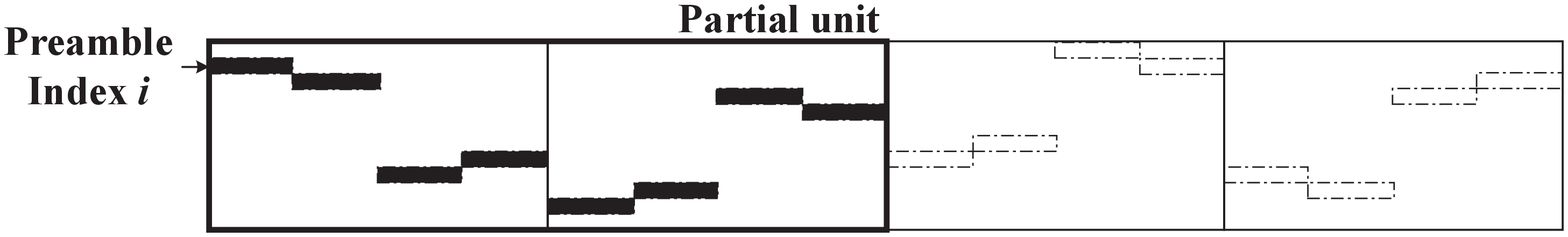}\label{fig:pt_2}}\\ 
    \vspace{-0.3cm}
    \subfigure[$L_p=16$ ($M_p=4$), $G=1$.]
    {\includegraphics[width=7.5cm]{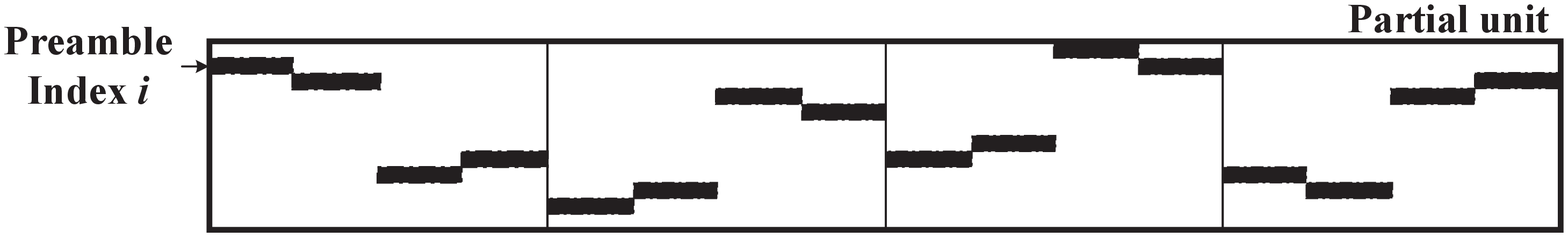}\label{fig:pt_3}}\\ 
    \vspace{-0.1cm}
    \caption[]{Configuration of partial preamble sequences when $L_b=16$ ($M_b$=4).
    }
    \label{fig:PPT_mechanism}
\vspace{-0.3cm}
\end{figure}

In the baseline ARP, when the amount of NPRACH resources is determined, the length of the preamble sequence is configured as $L_b=\nu \cdot M_b$, where $M_b$ represents the number of repetitions. However, in the proposed scheme, a PPS with a length of $L_p$ can be configured as $L_p=\nu \cdot M_p=\nu \cdot 2^{q_p}$, with $q_p=0,1,...,7$, where $M_p$ represents the number of repetitions in the PPSs. In this case, the eNodeB can configure $G$ non-overlapping PPSs, where $G=L_b/L_p$.
In Fig.~\ref{fig:PPT_mechanism}, we show three examples of the configuration of PPSs when $L_b=16$. Note that when $L_p=L_b$, the proposed ARP utilizes the assigned NPRACH in the same way as in the baseline ARP.

This partitioning of the preamble sequences may lead to degradation in the detection performance, while allowing the same preamble sequence to be reused up to $G$ times by multiple devices, thus reducing the occurrence of collisions. In essence, we establish a trade-off relationship between the occurrence of mis-detections and collisions. A decrease in the detection performance of the PPSs can be to some extent compensated by suitably increasing the PPS transmit power. 

\subsection{An Enhanced Access Reservation Protocol with a Partial Preamble Transmission Mechanism}\label{ARP_PPT}
The proposed ARP with the PPT mechanism mainly differs from the baseline NB-IoT ARP at the first two steps as follows:
\begin{itemize}
    \item {{\bf{Step 1}}: Each device randomly selects an index of preamble sequence
        among $N_\mathrm{P}$ preambles, and randomly selects \emph{a partial unit}
        among the $G$ available partial units. Each device transmits a PPS on the selected partial unit.
        }

    \item {{\bf{Step 2}}: The eNodeB determines which PPSs are received. The eNodeB
        accumulates the received power spread over each of the partial units, and compares
        it with the pre-defined detection threshold, $d_\mathrm{TH}$, at every partial
        unit\footnote{The number of detection events is increased by G times, however, the number of correlations per detection event remains the same, and, thus, this is not a severe burden for the eNodeB from the detection complexity perspective.}. If a certain PPS is detected, then the eNodeB transmits the RAR, which consists of an index of preamble, {\emph{an index of partial unit}}, a TA offset, and an uplink grant. Each device uses both the index of preamble and the index of partial unit to identify the destination of the RAR. }
\end{itemize}
%

\section{Performance Analysis}\label{CH3:Performance_Analysis}
We now mathematically characterize the detection and collision probabilities associated with the proposed ARP; and formulate an optimization problem where the objective is to maximize the ARP success probability.

\subsection{System Model} 
\label{sub:system_model}
We focus on a single NB-IoT cell, consisting of an eNodeB and IoT devices, which attempt to access the network through the ARP. Let $N_\mathrm{M}$ denote the
number of IoT devices which attempt the ARP in a single ARP session. Let $N_\mathrm{P}$ denote the number of preambles configured in a single NPRACH. We assume that each device performs an open-loop power control to compensate for the path loss. Thus, the channel between each device and the eNodeB can be modeled as a single-tap flat fading channel, where the channel coefficient follows a Rayleigh distribution, i.e., $h\sim\mathcal{CN}(0,1)$. For simplicity, we assume that the channel does not vary in a block of $\nu$ symbol groups, i.e., a single basic unit, but varies independently over the blocks.\footnote{Note that under the block fading channel model where $\nu$ is any positive integer, our mathematical analysis is still applicable.}

Let $\mathbf{y}^{k}$ denote the received signal of a tagged preamble sequence, which is simultaneously utilized by $k$ devices.
It can be expressed as $\mathbf{y}^{k}=\left[ y_{m,j,i}^{k} \right]$, for $m=1,...,M$, $j=1,...,\nu$, and $i=1,...,\xi$, where $y_{m,j,i}^{k}$ represents the $i$-th received symbol in the $j$-th symbol group at
the $m$-th repetition. $y_{m,j,i}^{k}$ can be represented as:
\begin{equation}
    y_{m,j,i}^{k}  = \sum\nolimits_{k' = 1}^{k} {\sqrt P  \cdot h_{m,j,i}^{k'}  \cdot x_{m,j,i} + w_{m,j,i}},
\end{equation}
where $P$, $h_{m,j,i}^{k'}$, $x_{m,j,i}$, and $w_{m,j,i}$ represent the received power per symbol at the eNodeB, the channel coefficient between the eNodeB and the $k'$-th device among $k$ devices which use the tagged preamble sequence, the $i$-th transmitted symbol in the $j$-th symbol group at the $m$-th repetition, and the Gaussian noise with \emph{zero} mean and variance of $2\sigma^2$, respectively.

\subsection{Normalized Received power}
To be able to perform the decision whether a certain PPS has been transmitted or not, the eNodeB needs to accumulate the received power of the PPS spread over the multiple symbol groups corresponding to the sequence. \emph{The normalized received power} of \emph{a tagged PPS}, which is transmitted simultaneously by $k$ IoT devices, $J_n^k$ is represented as:
\begin{equation}
    J_n^k  \!=\! \frac{1}{M_p} \sum\limits_{m = 1}^{M_p} {\left| {R_m } \right|^2 }  \!=\! \frac{1}{M_p} \sum\limits_{m = 1}^{M_p} {\left| {\sum\limits_{j = 1}^\nu  {\sum\limits_{i = 1}^\xi  {r_{y_{m,j,i}^{k} x_{m,j,i}} } } } \right|^2 },
\end{equation}
where $r_{yx}$ represents the correlation between $y$ and $x$ given by $r_{y x}=y \cdot x^*$, where $(\cdot)^{*}$ denotes the complex conjugate.
The term $r_{y_{m,j,i}^{k} x_{m,j,i}}$ can be expressed as:
\begin{equation}
    r_{y_{m,j,i}^{k} x_{m,j,i}} = \sum\nolimits_{k' = 1}^{k} {\sqrt P  \cdot h_{m,j,i}^{k'} + \tilde{w}_{m,j,i}},
\end{equation}
where $\tilde{w}_{m,j,i}$ follows the same distribution as ${w}_{m,j,i}$, and, thus, $r_{y_{m,j,i}^{k} x_{m,j,i}} \sim \mathcal{CN}(0, 2(kP+1)\sigma^2)$~\cite{Kim2016ICC}.
Therefore,
\begin{equation}
    J_n^k \sim \Gamma\left( {M_p,\frac{M_p}{{2(kP(\nu \xi )^2  + \nu \xi )\sigma^2 }}} \right),
\end{equation}
where $\Gamma\left( \alpha, \beta\right)$ represents a gamma distribution with shape $\alpha$ and rate $\beta$.

\subsection{False-alarm and mis-detection probabilities}
A \emph{false-alarm} occurs when a PPS that was not transmitted by any device, is detected to be active at the eNodeB. The false-alarm probability,  $p_\mathrm{fa}$, is expressed as:
\begin{align}\label{eq_pfa}
    p_\mathrm{fa}  &= \Pr \left\{ {J_n^0 > d_\mathrm{TH}} \right\} \\
            &= 1 - F\left( {d_\mathrm{TH};M_p,\frac{M_p}{{2\nu \xi \sigma ^2 }}} \right) \nonumber\\
            &= 1 - \frac{1}{{\Gamma \left( M_p \right)}} \cdot \gamma \left( {M_p,\frac{{d_\mathrm{TH}\cdot M_p}}{{2\nu \xi \sigma ^2 }}} \right), \nonumber
\end{align}
where $F(x; \alpha, \beta)$ represents the cumulative distribution function (CDF) of a gamma distribution, $\Gamma\left( \alpha, \beta\right)$. Note that $F(x; \alpha, \beta)$ can be expressed as $F(x; \alpha, \beta)= \frac{{\gamma (\alpha ,\beta x)}}{{\Gamma (\alpha )}}$, where $\gamma (\alpha ,\beta x)$ and $\Gamma(\cdot)$ represent the lower incomplete gamma function and the gamma function, respectively.

A \emph{mis-detection} occurs when a desired PPS is not detected, and its probability, $p_\mathrm{md}$, can be expressed as:
\begin{align}\label{eq_pmd}
p_\mathrm{md}  &= \frac{1}{\delta }\sum\nolimits_{k = 1}^{N_\mathrm{M} } {p_k }  \cdot \Pr \left\{ {J_n^k  < d_\mathrm{TH} } \right\} \\
        &= \frac{1}{\delta }\sum\nolimits_{k = 1}^{N_\mathrm{M} } {p_k }  \cdot F\left( {d_\mathrm{TH} ;M_p ,\frac{M_p}{{(kP(\nu \xi )^2  + \nu \xi )2\sigma ^2 }}} \right) \nonumber\\
        &= \frac{1}{\delta }\sum\nolimits_{k = 1}^{N_\mathrm{M} } {\frac{{p_k }}{{\Gamma \left( {M_p } \right)}}\cdot\gamma \left( {M_p ,\frac{{d_\mathrm{TH} \cdot M_p }}{{(kP(\nu \xi )^2  + \nu \xi )2\sigma ^2 }}} \right)}, \nonumber
\end{align}
where $\delta$ represents a scaling factor which is defined as $\delta=1-p_0$, while $p_k$ represents the probability that $k$ IoT devices simultaneously utilize the same tagged PPS, which is modeled by a binomial distribution as:
\begin{equation}
    p_k  = \left( \begin{array}{l}
 N_\mathrm{M}  \\
 ~k \\
 \end{array} \right)\left( {\frac{1}{{N_\mathrm{P}  \cdot G}}} \right)^k \left( {1 - \frac{1}{{N_\mathrm{P}  \cdot G}}} \right)^{N_\mathrm{M}- k }.
\end{equation}
%

\subsection{Collision probability}

A collision occurs when two or more IoT devices select the same PPS. When a tagged IoT
device selects a PPS, the average collision probability of the tagged IoT device, $p_\mathrm{c}$, can be
expressed as:
\begin{equation}
    p_\mathrm{c} = 1 - \left(1 - \frac{1}{N_\mathrm{P}\cdot G} \right)^{N_\mathrm{M}-1}.
\end{equation}
%

\subsection{Optimal PPS configuration}
The occurrence of mis-detections and collisions affects the ARP performance
\footnote{The limited amount of available uplink/downlink resources can also affect the ARP performance, yet the same level of performance degradation would be observed in the baseline protocol. Therefore, in this letter, our focus is solely on the effects associated with the expansion of the contention space through the PPT mechanism.}.
We evaluate this performance through the ARP success probability, $p_\mathrm{s}$, which is expressed as:
\begin{equation}
    p_\mathrm{s}=(1-p_\mathrm{c})(1-p_\mathrm{md}),
\end{equation}
where both $p_\mathrm{c}$ and $p_\mathrm{md}$ are functions of $M_p$, and, thus, we can formulate an optimization problem to find the optimal configuration of the PPSs as follows:
\begin{equation}\label{eq:opt_problem}
    M_{p}^\star = \mathop { \mathrm{argmax} }\limits_{M_p} p_\mathrm{s}(M_p).
\end{equation}
The optimal value of $M_p$ can be found numerically, while assuming that $M_b$, $P$, $\nu$, $\xi$ and an estimate of the number of contending devices is available.

\section{Numerical Results}\label{CH4:Numerical_results}
We present the numerical results related to the detection performance and success probability of the proposed ARP.
We have performed simulations using Matlab and the system parameters used during the evaluation are listed in Table~\ref{tb:simulation_env}.
\begin{table}
\centering
\caption{Simulation parameters and values}\label{tb:simulation_env}
\vspace{-0.2cm}
\begin{tabular} {>{\centering}m{5.9cm}|c}
    \Xhline{1.8\arrayrulewidth}
    Parameters & Values  \\
    \Xhline{1.8\arrayrulewidth}
    $\xi$, $\nu$ & 5, 4\\
    Number of preambles ($N_\mathrm{P}$) & 12 \\
    Number of repetitions in baseline ARP ($M_b$) & $64$ \\
    Number of repetitions in proposed ARP ($M_p$) & $2^{q_p}$,$q_p=0,..,7$\\
    Number of devices per a single ARP session ($N_\mathrm{M}$) & 1 $\sim$ 10 \\
    Detection threshold ($d_\mathrm{TH}$) & -5 $\sim$ 15 dB \\
    SNR ($\rho$) & -10, -5 dB\\
    \Xhline{1.8\arrayrulewidth}
\end{tabular}
\vspace{-0.2cm}
\end{table}
We first evaluate the impact of the PPT mechanism on the detection performance metrics for a single device, i.e. $N_\mathrm{M}=1$.

\begin{figure}[t]
    \centering
    \includegraphics[width=8.5cm]{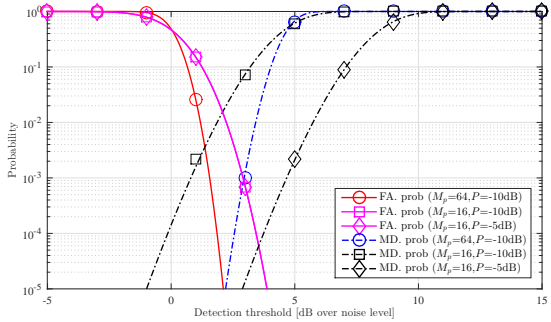} 
    \vspace{-0.3cm}
    \caption{Detection performance for varying the detection threshold, $d_\mathrm{TH}$. (Marker: Simulation, Line: Analysis)}
    \label{fig:detection_performance}
\vspace{-0.5cm}
\end{figure}

In Fig.~\ref{fig:detection_performance}, we depict $p_\mathrm{fa}$ and $p_\mathrm{md}$ for varying $d_\mathrm{TH}$ values.
Generally, the $p_\mathrm{fa}$ decreases and the $p_\mathrm{md}$ increases as the $d_\mathrm{TH}$ value increases when the $M_p$ value is given. Furthermore, a higher $M_p$ leads to a lower $p_\mathrm{fa}$ and $p_\mathrm{md}$. The former is due to noise averaging; the latter is due to a higher diversity gain. In the proposed ARP, only a fraction of a preamble sequence is transmitted by each device, therefore the eNodeB is not able to fully exploit the diversity associated with the original preamble sequence, and, thus, the detection performance degrades (i.e., a higher $p_\mathrm{md}$ is observed). However, if each device transmits the PPS with a higher transmit power, then the $p_\mathrm{md}$ can be decreased.
In practice, the preamble detector aims to provide a constant $p_\mathrm{fa}$, e.g., if we set a target $p_\mathrm{fa}$, i.e., $p_\mathrm{fa}^\mathrm{target}$, to $10^{-4}$, then the detection thresholds should be set to $1.86$dB and $3.44$dB, for $M_p=64$ and $M_p=16$, respectively.

\begin{figure}[t]
    \centering
    \includegraphics[width=8.5cm]{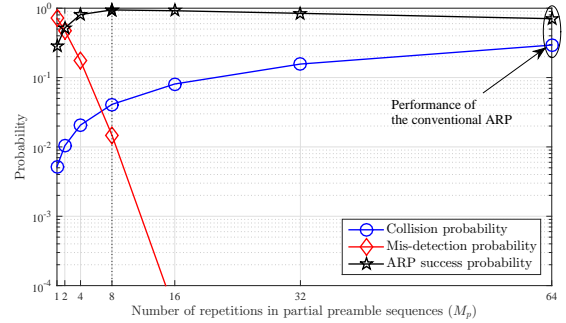} 
    \vspace{-0.3cm}
    \caption{Collision probability, mis-detection probability, and ARP success probability for varying the $M_p$ value of PPSs.}
    \label{fig:optimization_result}
\vspace{-0.2cm}
\end{figure}

Fig.~\ref{fig:optimization_result} shows the performance of the proposed ARP for varying the $M_p$ value of PPSs when $M_b\!=\!64$, $P\!=\!-5$dB, and $N_\mathrm{M}\!=\!5$. Decreasing the $M_p$ implies that the eNodeB can generate more contention resources from a single original preamble sequence while slightly sacrificing the detection performance. As the $M_p$ increases, the $p_\mathrm{md}$ decreases, however, the $p_\mathrm{c}$ increases. When the $M_p$ is equal to the $M_b$, the performance becomes the same as that of the baseline ARP. Due to this trade-off relationship according to how to utilize the given NPRACH resources, there exists an optimal point for maximizing the ARP success probability.

Table~\ref{tb:optimization_solution} shows the solutions of the optimization problem in Eq. (\ref{eq:opt_problem}) for varying $N_\mathrm{M}$ and $P$ when the parameters are given as $N_\mathrm{P}=12$, $M_b=64$, and $p_\mathrm{fa}^\mathrm{target}=10^{-4}$. The baseline ARP can guarantee extremely low mis-detection probabilities, since a sufficient number of repetitions are used to improve the detection performance. As a result, the ARP success probability is affected mostly by the collision probability.
On the other hand, in the proposed ARP, we can adjust the configuration of the PPSs. Therefore, when the system load is light, then it mitigates the mis-detection probability. However, when the system load becomes heavy, it mitigates the collision probability even though the detection performance degrades, and, thus, the ARP success probability can be drastically improved, compared to that of the baseline ARP. Note that if we consider more constraints on either $p_\mathrm{c}$ or $p_\mathrm{md}$ then the $M_p^{\star}$ may change.
\begin{table}[t]
\caption{The solutions of the optimization problem}
\label{tb:optimization_solution}
\vspace{-0.7cm}
\begin{center}
\setlength\tabcolsep{2pt}
\begin{tabular}{c|c||c|c|c|c||c|c|c|c}
\Xhline{2.0\arrayrulewidth}
\multicolumn{10}{c}{ $M_b=64$, $N_\mathrm{P}=12$, $p_\mathrm{fa}^\mathrm{target}=10^{-4}$} \\
\Xhline{2.0\arrayrulewidth}
\multicolumn{2}{c||}{Parameters} & \multicolumn{4}{c||}{Conventional ARP} & \multicolumn{4}{c}{Proposed ARP}\\
\Xhline{1.5\arrayrulewidth}
$N_\mathrm{M}$ & $P(\mathrm{dB})$ & $M_{b}$ & $p_\mathrm{c}(\%)$ & $p_\mathrm{md}$ & $p_\mathrm{s}(\%)$ & $M_{p}^{\star}$ & $p_\mathrm{c}(\%)$ & $p_\mathrm{md}$ & $p_\mathrm{s}(\%)$ \\ \hline
{\multirow{2}{*}{1}} & $-5$  & {\multirow{8}{*}{64}} & {\multirow{2}{*}{0}} & 0 & 100 & 64 & 0 & 0 & 100 \\
                     & $-10$ &  &  & $8.41e^{-7}$ & 100 & 64 & 0 & $8.41e^{-7}$ & 100 \\ \cline{1-2} \cline{4-10}
{\multirow{2}{*}{2}} & $-5$  &  & {\multirow{2}{*}{8.3}} & 0 & 91.7 & 16 & 2.08 & $4.46e^{-6}$ & 97.9 \\
                    & $-10$ &  &  & $8.04e^{-7}$ & 91.7 & 32 & 4.2 & $4.34e^{-3}$ & 95.4\\ \cline{1-2} \cline{4-10}
{\multirow{2}{*}{5}} & $-5$  &  & {\multirow{2}{*}{29.4}} & 0 & 70.6 & 8 & 4.10 & $1.48e^{-2}$ & 94.5 \\
                    & $-10$  &  &  & $7.01e^{-7}$ & 70.6 & 32 & 15.7 & $4.06e^{-3}$ & 84.0 \\ \cline{1-2} \cline{4-10}
{\multirow{2}{*}{10}} & $-5$  &  & {\multirow{2}{*}{54.3}} & 0 & 45.7 & 8 & 8.99 & $1.44e^{-2}$ & 89.7\\
                     & $-10$ &  &  & $5.51e^{-7}$ & 45.7 & 16 & 17.3 & $1.26e^{-1}$ & 72.3\\
\Xhline{2.0\arrayrulewidth}
\end{tabular}
\end{center}
\vspace{-0.6cm}
\end{table}

\vspace{-0.2cm}
\section{Conclusions} \label{conclusions}
We proposed an enhanced access reservation protocol (ARP) with a partial preamble transmission (PPT) mechanism for NB-IoT systems. The proposed ARP can mitigate the collision probability while slightly sacrificing the detection performance. We mathematically analyzed our proposed ARP in terms of the false alarm and mis-detection probabilities, and collision probability. We also investigated the trade-off relationship between the mis-detection probability and the collision probability, and found an optimal resource utilization strategy according to the system loads. Through extensive simulations, we verified that the proposed ARP outperforms the conventional NB-IoT ARP.



\vspace{-0.2cm}


\end{document}